\documentclass[twocolumn,aps,prl,10pt,showpacs]{revtex4-1}
\usepackage{bm}
\usepackage{amsfonts}
\usepackage{amssymb}
\usepackage{amsmath}
\usepackage{graphicx}
\begin{document}

\title{Relativistic Electron Wave Packets Carrying Angular Momentum}
\author{Iwo Bialynicki-Birula}\email{birula@cft.edu.pl}
\affiliation{Center for Theoretical Physics, Polish Academy of Sciences\\
Aleja Lotnik\'ow 32/46, 02-668 Warsaw, Poland}
\author{Zofia Bialynicka-Birula}
\affiliation{Institute of Physics, Polish Academy of Sciences\\
Aleja Lotnik\'ow 32/46, 02-668 Warsaw, Poland}

\begin{abstract}

 There are important differences between the nonrelativistic and relativistic description of electron beams. In the relativistic case the orbital angular momentum quantum number cannot be used to specify the wave functions and the structure of vortex lines in these two descriptions is completely different. We introduce analytic solutions of the Dirac equation in the form of exponential wave packets and we argue that they properly describe relativistic electron beams carrying angular momentum.
\end{abstract}
\maketitle
{\em Introduction.}---Recent advances in experiments with relativistic (100-300 keV) electron beams \cite{tonom,ver,uch,grillo,beche,hay,boyd1,boyd,saitoh} carrying orbital angular momentum call for a mathematical description based on the Dirac equation. The generally used Schr\"odinger equation gives an inadequate description because the differences between the nonrelativistic and relativistic wave functions are essential. It is not only the problem of relativistic corrections, which for 300 keV electrons may amount to about 60\%. A more important difference is in the use of the orbital angular momentum quantum number $l$ in the description of electronic states. In the nonrelativistic case both the orbital angular momentum and the total angular momentum are separately conserved while in the relativistic case only the total angular momentum has this property. This has already been pointed out by Dirac who in his first paper on the theory of the electron wrote ''This makes a difference between the present theory and the previous spinning electron theory, in which ${\bf m}^2$ is constant.'' Directly related to the problem of orbital angular momentum is a different structure of vortex lines in the two cases.

The nonrelativistic wave function in free space is simply a product of the coordinate part and the spin part; the orbital angular momentum and the spin are separately conserved. In the relativistic theory, even in free space, the spin is coupled to the orbital angular momentum and only the total angular momentum is conserved. As a consequence, there are no acceptable solutions of the Dirac equation that are eigenstates of the orbital angular momentum.

The proof of this assertion starts with the Dirac equation $i\hbar\partial_t\Psi=H\Psi$ and we assume that $L_z\Psi=\hbar l\Psi$. By multiplying both sides of the Dirac equation first by $\hbar l$, then by $L_z$, and subtracting the two equations one obtains $[L_z,H]\Psi=0$. Obviously, also $[L_z,H]^2\Psi=0$. Hence $(p_x^2+p_y^2)\Psi=0$. This means that $\Psi$ is a solution of the Laplace equation in 2D. However, in free space the solutions of this equation are unacceptable because they are all unbounded. In particular, the solutions with given $l$ behave as $\rho^l$ or $\rho^{-l}$.

There is also another important difference between the nonrelativistic and relativistic description. In the nonrelativistic case the speed of light does not appear. As a result, there is no intrinsic length scale. The relativistic length---Compton wavelength of the electron---plays a crucial role in our exponential solutions because it determines the asymptotic behavior of the wave packet far from the center.

The analysis of experiments in Refs.~\cite{tonom,ver,uch,grillo,beche,hay,boyd1,boyd,saitoh} is based on the solutions of the Schr\"odinger equation in the form of Laguerre-Gauss (LG) wave packets. The same nonrelativistic wave packets were the subject of theoretical analysis in Refs.~\cite{sclas,karimi,inmag,cons}. In principle, these solutions can be extended to the relativistic domain. However, they are not a good representation of the experimentally produced beams of relativistic electrons because, as we show below, they contain counterpropagating components. There is a solution of the Dirac equation in the form of Bessel functions \cite{bessel0,bessel} that does not have such components. However, Bessel functions are unphysical since they carry infinite energy. In this Letter we introduce solutions of the Dirac equation---exponential wave packets---that do not have these shortcomings. Like the Bessel and LG wave packets, the exponential wave packets are eigenfunctions of the {\em total angular momentum} in the direction of propagation. We believe that exponential wave packets are suitable representations of the experimentally studied beams of relativistic electrons endowed with angular momentum. In what follows we present a straightforward general procedure for constructing solutions of the Dirac equation and we apply this procedure to the three cases of interest.

{\em Construction of the solutions of the Dirac equation from scalar functions.}---Construction of the solutions of the Dirac equation is greatly simplified if one starts with solutions of the scalar Klein-Gordon (KG) equation. In order to generate a general solution of the Dirac equation one would need two such scalar functions. We shall be interested in the wave packets forming a beam of electrons with a given projection of the total angular momentum on the direction of propagation. In this case it is sufficient to use only one scalar function. Our construction (we will call it KG$\to$D) proceeds as follows. Let $f$ be a solution of the KG equation,
\begin{align}\label{kg}
\left[1/c^2\partial_t^2-\bigtriangleup + (mc/\hbar)^2\right]f=0.
\end{align}
Next, we form a two component spinor $\phi=(f,0)$ and act on this spinor with the matrix $i\lambdabar\mathcal D$,
\begin{align}\label{d}
\mathcal D=\left[\begin{array}{cc}1/c\,\partial_t+\partial_z&
\partial_x-i\partial_y\\\partial_x+i\partial_y&1/c\,\partial_t-\partial_z
\end{array}\right],
\end{align}
to define a second spinor $\chi$,
\begin{align}\label{dir1}
\chi=i\lambdabar{\mathcal D}\,\phi,
\end{align}
where $\lambdabar=\hbar/mc$ is the electron Compton wavelength. In turn, acting on $\chi$ with the conjugate matrix $\tilde{\mathcal D}$,
\begin{align}\label{dc}
\tilde{\mathcal D}=\left[\begin{array}{cc}1/c\,\partial_t-\partial_z&
-\partial_x+i\partial_y\\-\partial_x-i\partial_y&1/c\,\partial_t+\partial_z
\end{array}\right],
\end{align}
with the use of the KG equation for $\phi$, one obtains
\begin{align}\label{dir2}
\tilde{\mathcal D}\chi=-(i/\lambdabar)\phi.
\end{align}
Equations (\ref{dir1}) and (\ref{dir2}) can be written in a fully symmetric form
\begin{align}\label{dir3}
i\hbar{\mathcal D}\,\phi=mc\chi\\
i\hbar\tilde{\mathcal D}\,\chi=mc\phi.
\end{align}
This pair of equations is the Dirac equation for the bispinor $\Psi=(\phi,\chi)$,
\begin{align}\label{dir}
(i\hbar\gamma^\mu\partial_\mu-mc)\Psi=0,
\end{align}
written in the spinorial basis of $\gamma$ matrices,
\begin{align}\label{gamma}
\gamma^0=\left[\begin{array}{cc}0&1\\1&0\end{array}\right],\quad
\gamma^i=\left[\begin{array}{cc}0&-\sigma_i\\\sigma_i&0\end{array}\right].
\end{align}
Starting from different solutions of the KG equations, we shall now derive, by the KG$\to$D procedure, several solutions of the Dirac equation.

{\em Bessel wave packets of Dirac electrons.}---The Bessel solution of the Dirac equation is obtained by applying the KG$\to$D procedure to the scalar function $f_B^l$:
\begin{align}\label{fb}
f_{\rm B}^l(\rho,\varphi,z,t)=e^{-i(Et-p_zz)/\hbar}e^{il\varphi}J_l(p_\perp\rho/\hbar),
\end{align}
where $p_\perp=\sqrt{(E/c)^2-(mc)^2-p_z^2}$ is the transverse component of the momentum. In terms of spinors $\phi$ and $\chi$ this solution reads
\begin{align}\label{bes}
\phi=\left[\begin{array}{c}f_{\rm B}^l\\0\end{array}\right],\quad \chi=\frac{1}{mc}\left[\begin{array}{c}
(E/c-p_z)f_{\rm B}^l\\-ip_\perp f_{\rm B}^{l+1}\end{array}\right].
\end{align}

The Bessel solution of the Dirac equation $\Psi_B$ is labeled by three quantum numbers: the energy $E$, the momentum in the direction of propagation $p_z$, and the projection $\hbar l$ of the orbital angular momentum on the direction of propagation. The bispinor $\Psi_B$ is not an eigenfunction of $L_z$ because it contains parts with $l$ and $l+1$. However, it is an eigenfunction of the {\em total} angular momentum in the $z$ direction, $J_z=xp_y-yp_x+S_z$, belonging to the eigenvalue $\hbar(l+1/2)$. Should we have chosen $\phi=(0,f_B^l)$ and modified $\chi$ accordingly,
\begin{align}\label{bes1}
\phi=\left[\begin{array}{c}0\\f_{\rm B}^l\end{array}\right],\quad \chi=\frac{1}{mc}\left[\begin{array}{c}
ip_\perp f_{\rm B}^{l-1}\\
(E/c-p_z)f_{\rm B}^l\end{array}\right],
\end{align}
we would obtain the solution with the eigenvalue of $J_z$ equal to $\hbar(l-1/2)$.  The solutions (\ref{bes}) and (\ref{bes1}) illustrate the statement in the introduction that Dirac bispinors cannot be eigenfunctions of $L_z$. Indeed,  Eq.~(\ref{bes}) contains parts with $l$ and with $l+1$ while Eq.~(\ref{bes1}) contains parts with $l$ and with $l-1$. Even though pure Bessel solutions are not realistic because they carry infinite energy, we shall use them as very convenient building blocks, as we have done before for optical beams \cite{bb,bb1}.

{\em Laguerre-Gauss wave packets of Dirac electrons.}---The LG solution of the Schr\"odinger equation, frequently mentioned in the context of electron beams \cite{boyd,uch,ver,grillo,sclas,cons,karimi,inmag} (although never written down explicitly), has the form
\begin{align}\label{slg}
\psi_{\rm LG}&(\rho,\varphi,z,t)=\exp\left(-i p_z^2t/2m\hbar \right)\exp\left(ip_zz/\hbar\right)
\nonumber\\&\times\frac{\rho^{|l|}e^{il\varphi}}{a(t)^{n+|l|+1}}
\exp\left(-\frac{\rho^2}{a(t)}\right)L_n^{|l|}\left(\frac{\rho^2}{a(t)}\right),
\end{align}
where $a(t)=w^2+2i\hbar t/m$. The parameter $w$ controls the width of the wave packet. This solution of the nonrelativistic Schr\"odinger equation can be extended to the relativistic domain by choosing the scalar solution of the KG equation $f_{\rm LG}$ in the form
\begin{align}\label{flg}
f_{\rm LG}&(\rho,\varphi,z,t)=\exp\left(-i E t_-/2\hbar\right)\exp\left(-i\frac{m^2c^4}{2E\hbar}t_+\right)
\nonumber\\&\times\frac{\rho^{|l|}e^{il\varphi}}{a(t_+)^{n+|l|+1}}
\exp\left(-\frac{\rho^2}{a(t_+)}\right)L_n^{|l|}\left(\frac{\rho^2}{a(t_+)}\right),
\end{align}
where $a(t_+)=w^2+2i\hbar c^2t_+/E$ and $t_\pm=t\pm z/c$. In the nonrelativistic limit ($c\to\infty$), after setting $E=mc^2+cp_z$ and filtering out the rest mass oscillations $\exp(imc^2t/\hbar)$, one obtains back $\psi_{\rm LG}$.

The LG solution of the Dirac equation $\Psi_{\rm LG}$ is obtained by applying the KG$\to$D procedure to $f_{\rm LG}$. This solution does not describe correctly relativistic electron beams because in addition to the dependence on $t-z/c$ it depends also on $t+z/c$. The dependence on $t_+$ invalidates the use of the LG solutions for relativistic electrons because it simply means that in such a beam there are also electrons propagating in the opposite direction. We propose to replace the LG solutions of the Dirac equation by the exponential solutions described below.

\begin{figure}
\includegraphics[scale=0.9]{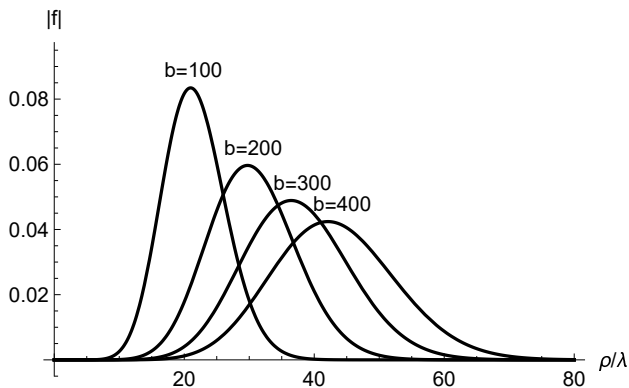}
\caption{Normalized modulus $|\psi|$ of the wave function (\ref{exp}) as a function of $\rho/\lambdabar$ plotted for $l=10$ and different values of the parameter $b$.}\label{fig1}
\end{figure}

{\em The exponential wave packets of Dirac electrons.}---The exponential solutions of the Dirac equation are obtained by applying the KG$\to$D procedure to the following scalar solution of the KG equation:
\begin{align}\label{exp}
f_{\rm Exp}&(\rho,\varphi,z,t)=e^{ip_zz/\hbar}e^{il\varphi}\nonumber\\
&\times\frac{e^{-bh(\rho,t)}}{h(\rho,t)}
\left(\frac{q\rho}{h(\rho,t)+1+iqct}\right)^{|l|},
\end{align}
where
\begin{align}\label{h} h(\rho,t)=\sqrt{\left(1+iqct\right)^2+\left(q\rho\right)^2},
\end{align}
$b$ is a dimensionless parameter that controls the width of the wave packet, $q=\gamma/b\lambdabar$, and $\gamma=\sqrt{1+(p_z/mc)^2}=1/\sqrt{1-v_z^2/c^2}$ is the relativistic factor. For realistic beams the parameter $b$ must be very large to compensate for the smallness of the electron Compton wavelength ($\lambdabar=3.86\times 10^{-13}$m). The exponential solutions are labeled by two quantum numbers $p_z$ and $l$ and by the parameter $b$. In Fig.~\ref{fig1} we show that, indeed, the parameter $b$ determines the width of the wave packet. The beams described by the exponential solutions for large values of $\rho/\lambdabar$ fall off as $\exp(-\gamma\,\rho/\lambdabar)/\rho$. In contrast to relativistic LG solutions, this rate of decrease is not constant but it grows with increasing electron energy.

Applying the KG$\to$D procedure one obtains the exact solution $\Psi_{\rm Exp}$ of the Dirac equation with the spinors $\phi$ and $\chi$ built from $f_{\rm Exp}$,
\begin{align}\label{expd}
\phi=\left[\begin{array}{c}f_{\rm Exp}\\0\end{array}\right],\quad \chi=i\lambdabar\left[\begin{array}{c}
(1/c\,\partial_t+ip_z/\hbar)f_{\rm Exp}\\e^{i\varphi}(\partial_\rho-l/\rho)f_{\rm Exp}\end{array}\right].
\end{align}
This bispinor describes the state with the component $J_z$ of the total angular momentum equal to $\hbar(l+1/2)$. Like in the case of Bessel beams, the alternative choice $\phi=(0,f_{\rm Exp})$ produces the solution with the total angular momentum equal to $\hbar(l-1/2)$.

It is not obvious that our exponential wave packets, in contrast to the LG solution, have no counterpropagating components. This property will be proved by expanding $f_{\rm Exp}$ into Bessel solutions.

{\em Expansion of the exponential wave packets into Bessel solutions.}---Every scalar solution of the KG equation can be written as a superposition of Bessel solutions since they form a complete set. To obtain the exponential wave packet (\ref{exp}) it suffices to include only Bessel solutions (\ref{fb}) with fixed $l$ and $p_z$. Such a superposition has the following general form:
\begin{align}\label{exp0}
f_{\rm Exp}(\rho,\varphi,z,t)&=\int_{E_\parallel}^\infty dE\, e^{-i(Et-p_zz)/\hbar}e^{il\varphi}\nonumber\\
&\times g(E)J_l(\sqrt{E^2-E_\parallel^2}\,\rho/\hbar c),
\end{align}
where $E_{||}=\gamma mc^2$ is the energy associated with the momentum in the $z$ direction. The factor $e^{-i(Et-p_zz)/\hbar}$ in this formula guarantees that $f_{\rm Exp}$ describes positive energy solutions (electrons and not positrons) and that there are no counterpropagating components.

As we have shown in Ref.~\cite{bb1}, there are several choices of the spectral function $g(E)$ that allow for an analytic evaluation of the integral and lead to exponential wave packets. The simplest exponential wave packet (\ref{exp}) is obtained for
\begin{align}\label{spec}
g(E)=b\frac{e^{-b E/E_{||}}}{E_{||}}
\left(\frac{E-E_{||}}{E+E_{||}}\right)^{|l|/2}.
\end{align}
The integration over $E$ can be performed with the use of the formula 6.646.1 in Ref.~\cite{gr}.
\begin{figure}
\includegraphics[scale=0.9]{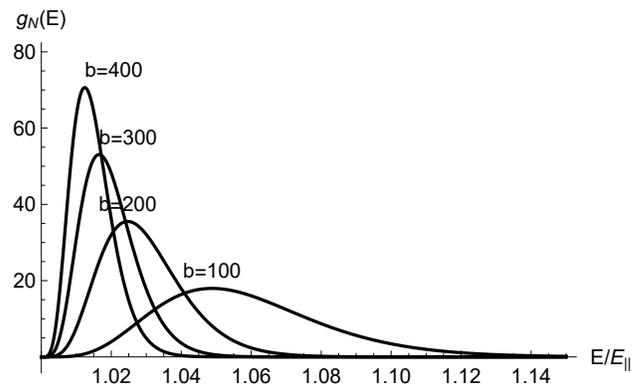}
\caption{Normalized spectral functions $g_N(E)$ plotted for $l=10$ and different values of the parameter $b$.}\label{fig2}
\end{figure}
The maximum of the spectral function (\ref{spec}) is located at $E=\gamma mc^2\sqrt{1+|l|/b}$. When $b$ increases, the spectral function tends to $\delta(E-E_{||})$. This is illustrated in Fig.~\ref{fig2} where the normalized spectral function $g_N(E)=g(E)/\int dE\,g(E)$ is plotted for different values of $b$. Therefore, the parameter $b$ controls the monochromaticity of the wave. For very large values of $b$ the exponential beam approaches the Bessel wave packet. However, this convergence is not uniform; it holds only for restricted values of $\rho$ and $t$. The character of the dependence on the parameter $b$, shown in Figs.~\ref{fig1} and \ref{fig2}, may be viewed as a manifestation of the uncertainty principle; the broader the wave packet in coordinate space, the sharper the spectrum.

{\em Nonrelativistic limit of the exponential wave packet.}---The nonrelativistic limit is obtained here in a more complicated way than in the case of the function $f_{\rm LG}$. Before evaluating the limit $c\to\infty$, in addition to filtering out the rest mass oscillations one must introduce now also a $c$-dependent rescaling of the parameter $b=\beta\,c^2$ and change the normalization of the wave function. The resulting solution of the Schr\"odinger equation is simply the nonrelativistic LG wave packet (\ref{slg}) with $n=0$,
\begin{align}\label{sexp}
\psi_{\rm LG}(\rho,\varphi,z,t)|_{n=0}&=\exp\left(-i p_z^2t/2m\hbar \right)\exp\left(ip_zz/\hbar\right)
\nonumber\\&\times\frac{\rho^{|l|}e^{il\varphi}}{a(t)^{|l|+1}}
\exp\left(-\frac{\rho^2}{a(t)}\right),
\end{align}
where $a(t)$ contains now the parameter $\beta$, namely $a(t)=2\beta\,(\hbar/m)^2+2it\hbar/m$.

{\em Vorticity and vortex lines}---In the nonrelativistic description there is no ambiguity in defining the velocity ${\bm v}$ of the probability flow,
\begin{align}\label{vel}
{\bm v}=\frac{\hbar}{m}\frac{{\rm Im}(\psi^*{\bm\nabla}\psi)}{\psi^*\psi}=\frac{\hbar}{m}{\bm\nabla}S.
\end{align}
Wave functions endowed with orbital angular momentum have a characteristic phase $S=l\varphi$. Owing to the singular character of $S$, the vorticity ${\bm w}={\bm\nabla}\times{\bm v}$ is localized on vortex lines that undergo intricate time evolution dictated by the Schr\"odinger equation \cite{bbs}.

In the relativistic theory the velocity ${\bm v}_D$ is obtained from the Dirac four-current $j^\mu={\bar\Psi}\gamma^\mu\Psi$,
\begin{align}\label{vel1}
{\bm v}_{\rm D}=c\frac{{\bar\Psi}{\bm\gamma}\Psi}{\Psi^\dagger\Psi}
=c\frac{\phi^\dagger{\bm\sigma}\phi-\chi^\dagger{\bm\sigma}\chi}
{\phi^\dagger\phi+\chi^\dagger\chi}.
\end{align}
\begin{figure}
\includegraphics[scale=0.7]{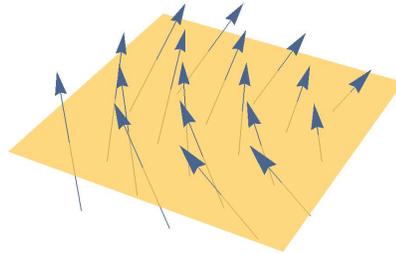}
\caption{Distribution of vorticity in the exponential beam.}\label{fig3}
\end{figure}
In contrast to the nonrelativistic case, the velocity is now not a gradient of a phase. As a result, the vorticity ${\bm w}_D={\bf\nabla}\times{\bf v}_D$ is not concentrated on vortex lines but it is continuously spread all over space (see Fig.~\ref{fig3}). We shall resolve the seeming discrepancy between the nonrelativistic and relativistic behavior of vorticity with the use of the Gordon decomposition \cite{gordon} of the Dirac current into the orbital part and the spin part, ${\bm j}={\bm j}_{\rm orb}+{\bm j}_{\rm spin}$,
\begin{subequations}\label{gordon}
\begin{align}
{\bm j}_{\rm orb}&=\frac{2\hbar}{m}\,{\rm Im}\left(\phi^\dagger{\bm\nabla}\chi\right),\\
{\bm j}_{\rm spin}&=\frac{\hbar}{m}\,\left[{\bm\nabla}\times
{\rm Re}\left(\phi^\dagger{\bm\sigma}\chi\right)
-\partial_t{\rm Re}\left(\phi^\dagger{\bm\sigma}\chi\right)\right].
\end{align}
\end{subequations}
Near the beam center the solutions of the Dirac equations studied here behave similarly and we use the Bessel solution (\ref{bes}) to make the calculations simple. In this case the leading part of the orbital velocity ${\bm v}_{\rm orb}={\bm j}_{\rm orb}/(\phi^\dagger\phi+\chi^\dagger\chi)$, in the limit $c\to\infty$, is
\begin{align}\label{vort}
{\bm v}_{\rm orb}\approx\frac{\hbar\,l}{m}
(-\frac{y}{\rho^2},\frac{x}{\rho^2},0).
\end{align}
This singular form of the velocity produces the vortex line along the $z$ axis, in full agreement with the nonrelativistic formula (\ref{vel}). The vorticity localized on the vortex line associated with the orbital velocity is, however, {\em exactly canceled} by the opposite vorticity associated with the spin part of the velocity. As a result, the vorticity in the wave packet described by the Dirac equation has no singularities; it is distributed continuously in space.

{\em Discussion.}--- The differences between the nonrelativistic and relativistic quantum theory of electrons are so substantial that they completely invalidate some conclusions based on the Schr\"odinger equation. The first difference is the nonexistence of relativistic electron wave packets with fixed orbital angular momentum. The second difference is the presence of the relativistic length parameter---Compton wavelength---which determines the behavior at large distances from the center. The third difference is lack of freedom to manipulate separately orbital angular momentum and spin. Finally, there is an open problem of vortex lines in the relativistic case which boils down to the question: Which current, total (no vortex lines) or orbital (with vortex lines), is observed in experiments?

This research was partly supported by Polish National Science Center Grant No. 2012/07/B/ST1/03347.

\end{document}